\documentclass[epj-spec]{svjour}
\usepackage{graphics}
\begin{document}
\title{Statistical Relaxation in Closed Quantum Systems and the Van Hove-Limit}
%\subtitle{Do you have a subtitle?\\ If so, write it here}
\author{Christian Bartsch\inst{1}\fnmsep\thanks{\email{cbartsch@uos.de}} \and Pedro Vidal\inst{2}\fnmsep\thanks{\email{pedro@itp1.uni-stuttgart.de}}  }
\institute{Universtit\"at Osnabr\"uck, Fachbereich Physik, Barbarastra\ss e 7, D-49069 Osnabr\"uck \and Universit\"at Stuttgart, I Institut f\"ur Theoretische Physik, Pfaffenwaldring 57 // IV 70550 Stuttgart}
\abstract{
%Insert your abstract here.
We analyze the dynamics of occupation probabilities for a certain type of design models by the use of two different methods. On the one hand we present some numerical calculations for two concrete interactions which point out that the occurrence of statistical dynamics depends on the interaction structure. Furthermore we show an analytical derivation for an infinite system that yields statistical behaviour for the average over the whole ensemble of interactions in the Van Hove-limit.
} %end of abstract
\maketitle
\section{Introduction}
\label{intro}

The emergence of statistical behaviour from microscopic dynamics is of special interest for two reasons. Its existence is evident from countless experiments, nevertheless its explanation seems subtle. And if statistical dynamics are established, their description is much simpler than the description of microscopic dynamics. From first principles the dynamics of quantum systems are controlled by the Schr\"odinger equation. Nevertheless, it has been observed that statistical relaxation may appear in such systems for certain quantities under certain conditions. The dynamics of the quantities $P_{n}$ are called statistical if they are given by a master equation of the form
\begin{equation}
\frac{\partial}{\partial t}P_{n} = \sum_{m} R(m \rightarrow n)P_{m}-\sum_{m} R(n \rightarrow m)P_{n}\ .
\label{rate}
\end{equation}
This is a set of coupled rate equations whose solutions decay exponentially in time.

In this contribution the dynamics for a certain type of modular design model are approached from two different sides. On the one hand, we show numerical calculations of the time evolution of occupation probabilities for a finite size version of the model with two concrete types of interactions. Furthermore we will present an
analytical derivation that analyzes the dynamics of these variables for an average over all possible interactions in the Van Hove-limit and in the limit of an infinitely large system size, in the sense that the number of energy levels in the bands of the modules $N$ goes to $\infty$, whereas the number of modules $M$ is
always $2$, i.e., $M$ is still finite.
%which indicates that the dynamics of those variables are statistical for the average over all possible interactions in the limit of an
%infinitely large system size and in the Van Hove limit.....

The numerics are consistent with the analytical results for an interaction which is kept very general and represents the majority of all interactions. Nevertheless,  there can be completely exceptional interactions that yield substantially different dynamics.

%Your text comes here. Separate text sections with
\section{Numerical calculations}
\label{sec:1}

\subsection{Design model}
\label{model}

\begin{figure}[h!]
\begin{center}
\resizebox{0.55\columnwidth}{!}{
  \includegraphics{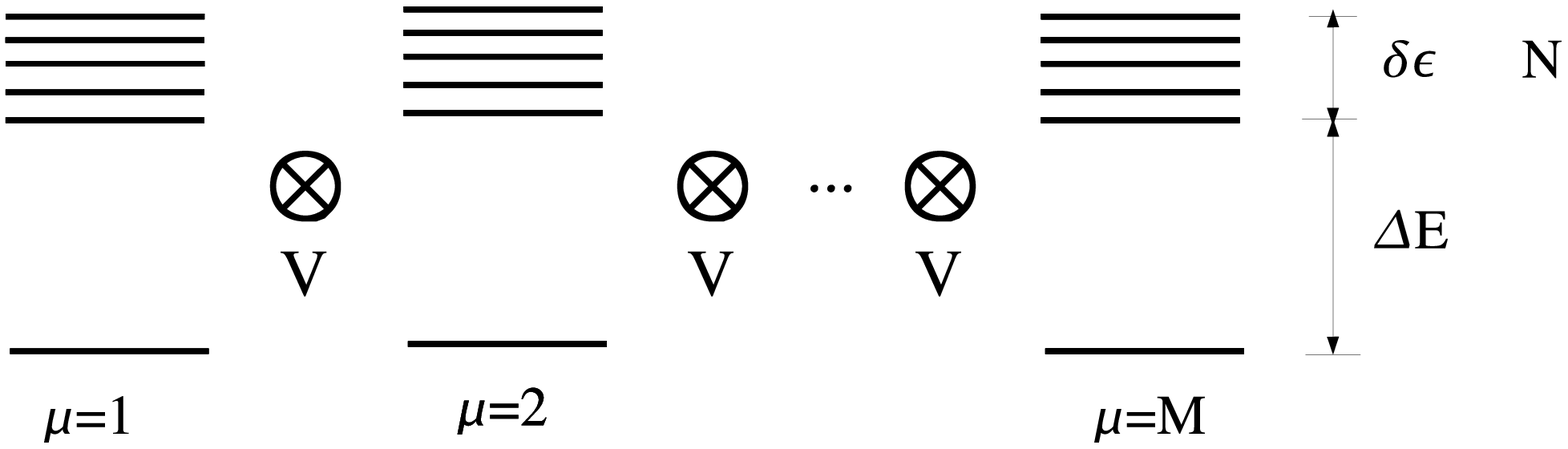} }
\caption{Sketch of the design model}
\label{designmodel}
\end{center}
\end{figure}

The model we investigate for the occurrence of statistical dynamics is a simple design model depicted in Fig.~\ref{designmodel} (see also \cite{buchquanttherm,fourierlaw,statrelschr,heatcond}).

It consists of $M$ one-dimensionally arranged identical modules that are coupled to the two neighbouring modules via the interaction $V$.
Each subunit consists of a single ground state and an energy band. These are separated by the band gap $\Delta E$. The model can be characterized by some rough
system parameters; $N$ is the number of levels in each band, $\delta\epsilon$ denotes the band width ($\delta\epsilon\ll\Delta E$) and $\lambda$ gives the average interaction strength. The Hamiltonian can be split into a local part $H_{0}$ and an interaction $V$ 
\begin{equation}
H=H_{0}+\lambda V\ .
\end{equation}
The interaction is chosen to be weak such that one can assume local energy conservation and restrict to the subspace of the states where one system is in the
excited state and all other systems are in the ground state. In this contribution only chains with two modules ($M=2$) are regarded. 
$V$ shall have an off-diagonal block form in the eigenbasis representation of $H_{0}$ in the relevant subspace, thus the Hamiltonian is given by
\begin{center}
     \begin{small}
     \begin{equation}
     H=
     \left(\begin{array}{ccccc|ccccc}
     \ddots & & & &0 & & & & \\
     & & & & & & & & &\\
     & &\Delta E+\frac{i}{N-1}\delta \epsilon & & & & &V & & \\
     & & & & & & &\\
     0& & & &\ddots & & & & &\\
     \hline
     & & & & &\ddots & & & &0\\
     & & & & & & & & &\\
     & &V^{\dagger} & & & & &\Delta E+\frac{j}{N-1}\delta \epsilon & &\\
     & & & & & & & & &\\
     & & & & &0 & & & &\ddots\\
     \end{array}\right)
     \label{hamil}
     \ .
     \end{equation}
     \end{small}
\end{center}
The relevant space is divided into two subspaces each containing the states where the $\mu$-th subsystem is in the excited state ($\mu=1,2$). $V$ is normalized by
\begin{equation}
\frac{{\rm Tr}\{ V^{2}\}}{2N^{2}}=1\ .
\end{equation}
One can now define the occupation probability in the $\mu$-th subsystem $P_{\mu}$.  
%which can be written as the expectation value of the projector on the $\mu$-th
According to the predictions of the Hilbert Space Average Method (HAM) in second order (see contribution Breuer/Gemmer) the time evolution of the $P_{\mu}$ is controlled by coupled rate equations which means that the dynamics are statistical
\begin{eqnarray}
\frac{dP_{1}}{dt}=RP_{2}-RP_{1}\ ,\nonumber\\
\frac{dP_{2}}{dt}=RP_{1}-RP_{2}\ ,
\label{rateeq}
\end{eqnarray}   
\begin{equation}
{\rm with} \hspace{1.5cm} R=2\pi\lambda g \hspace{1.5cm} {\rm and} \hspace{1.5cm} g=\frac{N}{\delta\epsilon}\ .
\end{equation}
The rate is completely determined by the system parameters and is the same rate that appears in the context of Fermi's Golden Rule. In fact, HAM uses some
kind of stepwise iteration of Fermi's Golden Rule. One also obtains some necessary conditions for the system parameters,
\begin{equation}
K_{1}=\lambda^{2}\frac{N}{\delta\epsilon^{2}}\ll 1\ ,
\label{kriterium1}
\end{equation}
\begin{equation}
K_{2}=2\lambda\frac{N}{\delta\epsilon}\geq 1\ .
\label{kriterium2}
\end{equation} 
A more detailed derivation can be found in \cite{buchquanttherm,fourierlaw,statrelschr,heatcond}. 

In particular, these results do not depend on the concrete realization of $V$. So, in principle, they should be valid for any possible interaction structure.

\subsection{Numerical calculations for two concrete interactions}
\label{numerics}

In this paragraph we are going to compare the dynamics of the $P_{\mu}$ obtained from the rate equations (\ref{rateeq}) with the numerical solution of
the Schr\"odinger equation for two specific interactions. The system parameters are adjusted in a way that the criteria (\ref{kriterium1}) and (\ref{kriterium2}) are well fulfilled. 

For the first interaction the matrix elements of $V$ in the eigenbasis representation of $H_{0}$ are chosen as random Gaussian distributed complex numbers.
For the second interaction all matrix elements of $V$ are set to be equal, say, $V_{ij}=1$. The random interaction possesses no structure at all, whereas
the opposite holds true for the constant interaction.
% is very strongly structured.

\begin{figure}[h!] 
\begin{minipage}[t]{0.47\textwidth} 
\begin{center} 
\resizebox{\columnwidth}{!}{
\includegraphics{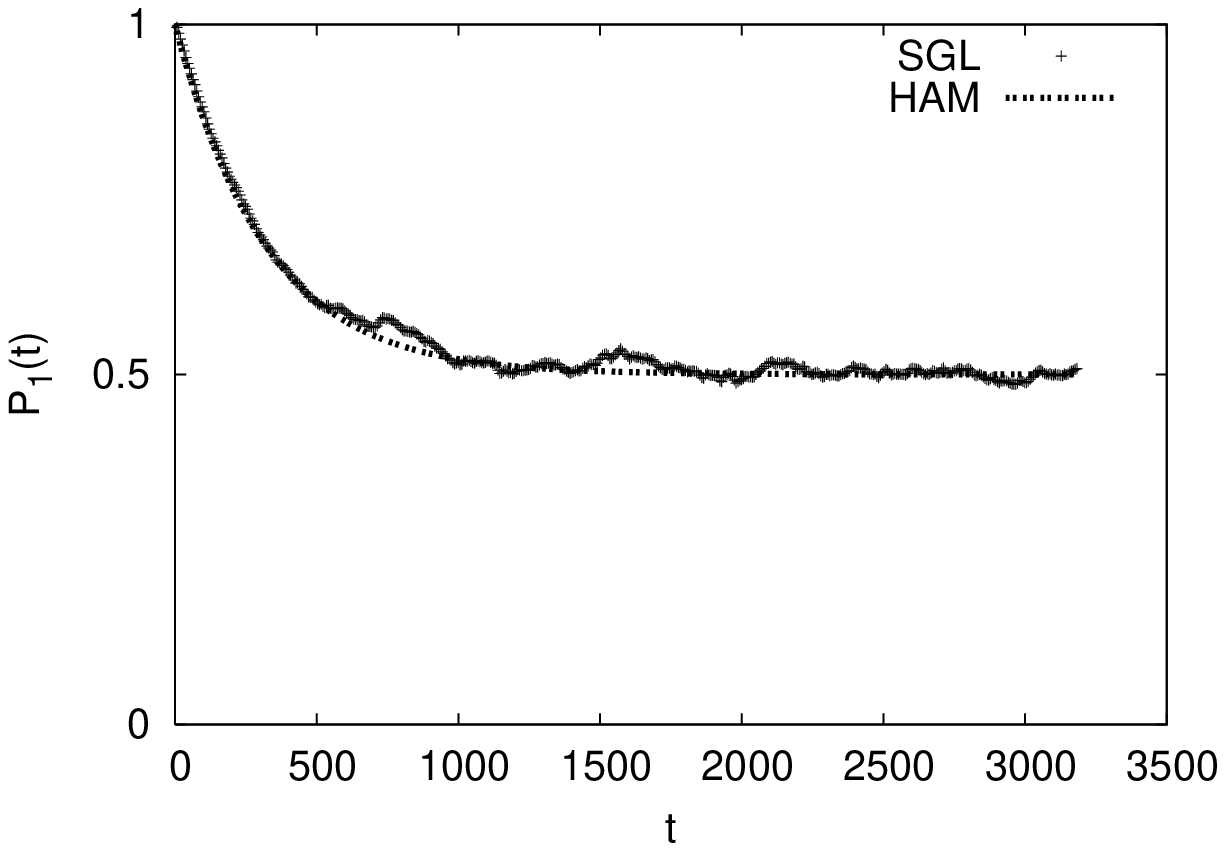} }
\caption{Dynamics of $P_{1}(t)$ for a completely random interaction, parameters: 
 $N=500$, $\delta\epsilon=0.5$, $\lambda=0.0005$.} 
\label{randWW}
\end{center} 
\end{minipage}\hfill 
\begin{minipage}[t]{0.47\textwidth} 
\begin{center} 
\resizebox{\columnwidth}{!}{
\includegraphics{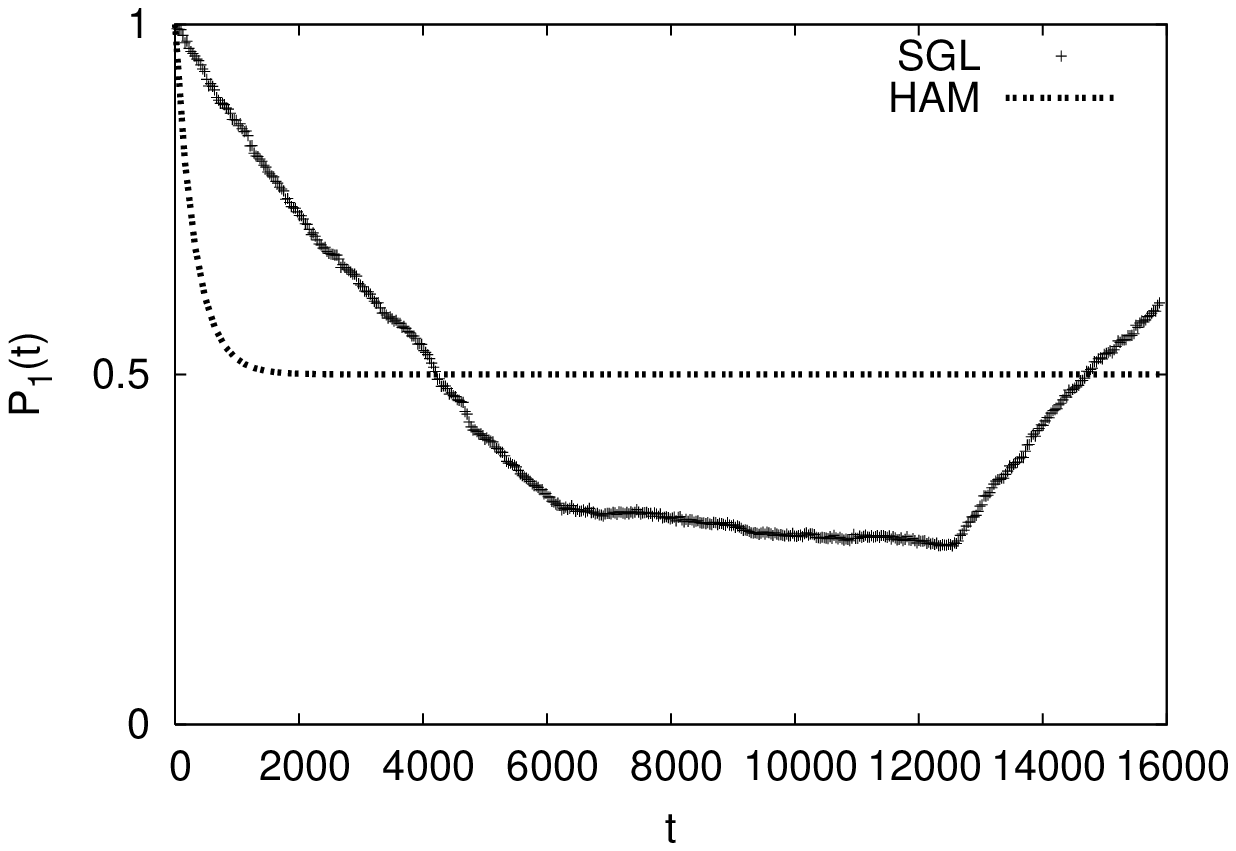} }
\caption{Dynamics of $P_{1}(t)$ for a constant interaction with $V_{ij}=1$, parameters:
 $N=500$, $\delta\epsilon=0.5$, $\lambda=0.0005$.} 
\label{constWW}
\end{center} 
\end{minipage} 
\end{figure}

Figure~\ref{randWW} and Fig.~\ref{constWW} show the time evolution of $P_{1}(t)$ for both interactions. The initial state is selected to have $P_{1}(0)=1$.

One finds a very good agreement between both curves for the random interaction. Therefore the rate equations predicted by HAM are valid and the dynamics are
statistical. For the constant interaction there is no correspondence at all, so one finds no statistical relaxation in this case, although the system parameters, which are the same for both interactions, fulfill the criteria (\ref{kriterium1}) and (\ref{kriterium2}). 
Referring to Sec.~\ref{sec:2} the random interaction represents the average over all possible interactions very well, whereas the constant interaction
demonstrates that there can be complete exceptions from the average.
% is a complete exception (shows that complete exceptions from the average exist).
It becomes obvious that the concrete structure of the interaction can be indeed significant for the occurrence of statistical relaxation. 
%This can be caused by the
%fact that the influence of higher orders may vary for different interactions.

\begin{figure}[h!] 
\begin{minipage}[t]{0.47\textwidth} 
\begin{center} 
\resizebox{\columnwidth}{!}{
\includegraphics{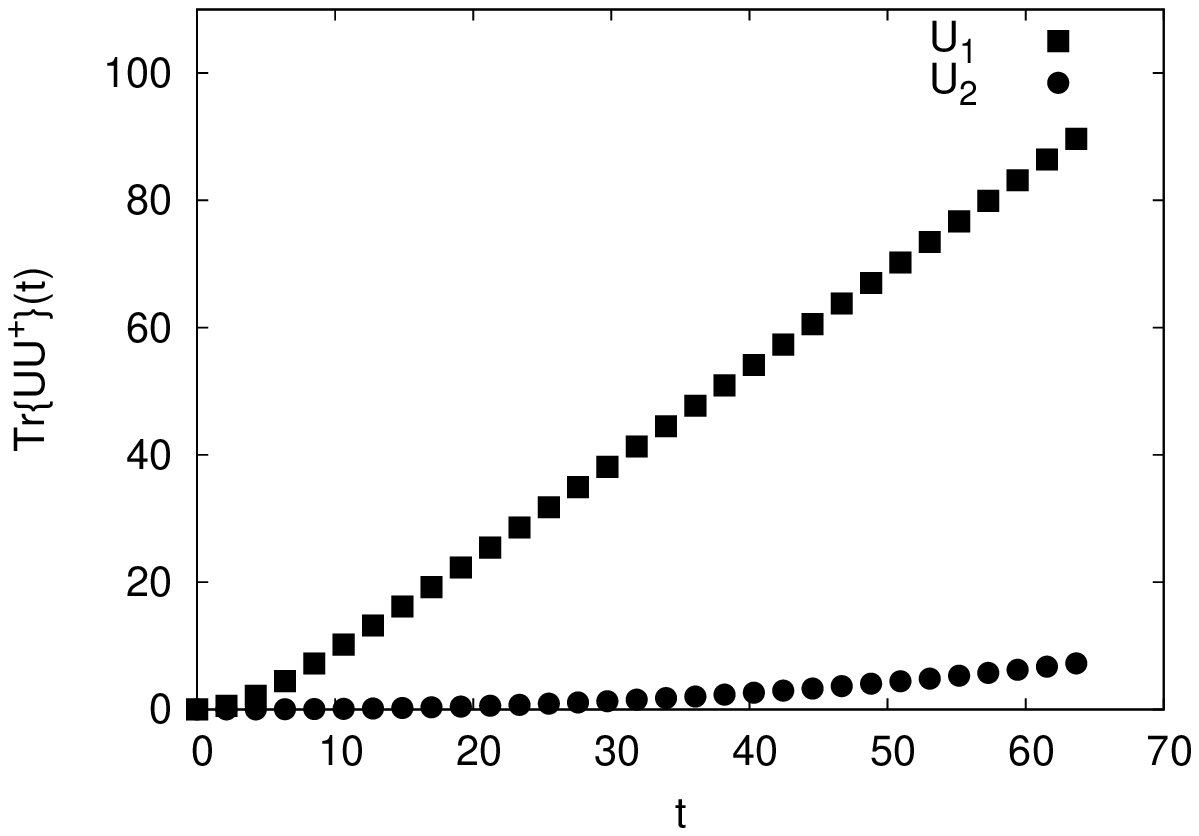} }
\caption{Dyson terms for a completely random interaction, parameters: 
 $N=500$, $\delta\epsilon=0.5$, $\lambda=0.0005$.} 
\label{dysonrandWW}
\end{center} 
\end{minipage}\hfill 
\begin{minipage}[t]{0.47\textwidth}
\begin{center} 
\resizebox{\columnwidth}{!}{
\includegraphics{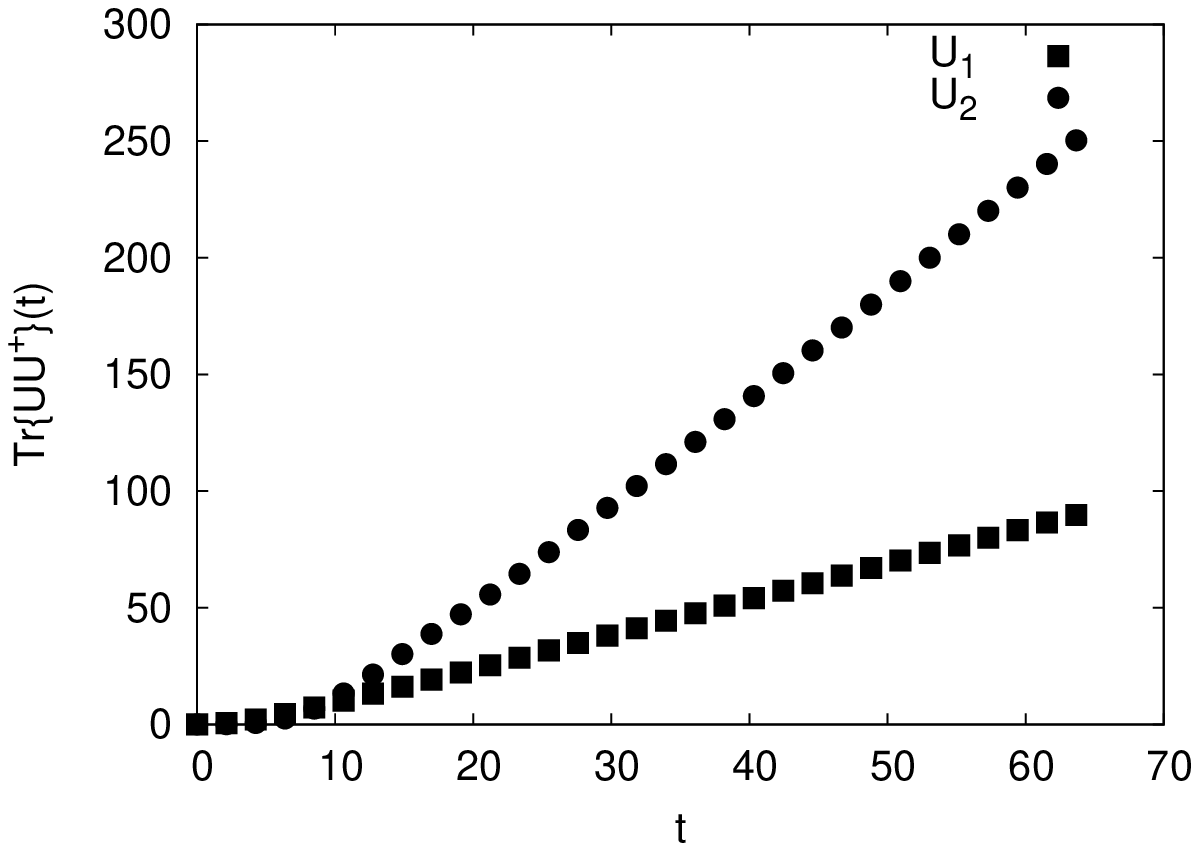} }
\caption{Dyson terms for a constant interaction with $V_{ij}=1$, parameters:
 $N=500$, $\delta\epsilon=0.5$, $\lambda=0.0005$.} 
\label{dysonconstWW}
\end{center} 
\end{minipage} 
\end{figure}

The HAM calculation, as well as Fermi's Golden Rule, relies on second order perturbation theory, i.e., the time evolution operator is expressed by the Dyson series
in second order truncation 
%\begin{equation*}
%\hat{D}(\tau)\approx \hat{1}-\frac{i}{\hbar}\hat{U}_{1}(\tau)-\frac{1}{\hbar^{2}}\hat{U}_{2}(\tau)
%\end{equation*}
\begin{equation}
\vert\psi (\tau)\rangle=(\hat{1}-\frac{i}{\hbar}U_{1}(\tau)-\frac{1}{\hbar^{2}}U_{2}(\tau)+...)\vert\psi(0)\rangle
\end{equation}
with the first and second order terms
\begin{eqnarray}
U_{1}(\tau)&=& \int_{0}^{\tau}d\tau 'V(\tau ')\ ,\\
U_{2}(\tau)&=& \int_{0}^{\tau}d\tau ' \int_{0}^{\tau '}d\tau ''V(\tau ')V(\tau '')\ .
\end{eqnarray}
If this expansion is justified, all higher orders, including $U_{2}(\tau)$, must be small compared to $U_{1}(\tau)$ in the relevant time regime. 
We use $\rm{Tr}\{ UU^{\dagger}\}$ to measure the size of the respective contributions. 
HAM produces a "best guess" for the time evolution of the $P_{\mu}$ for a short time step ($P_{\mu}(t+\tau)$) on the basis of $P_{\mu}(t)$ by using the
appropriate truncation of the Dyson series,
\begin{equation}
P_{\mu}(t+\tau)-P_{\mu}(t)\approx f(\tau)(P_{\mu-1}(t)+P_{\mu+1}(t)-2P_{\mu}(t))\ ,
\label{ststep}
\end{equation}
where $f(\tau)$ corresponds to a double time integral over the autocorrelation function of $V$ (see also \cite{fourierlaw}). One usually expects that this autocorrelation function has 
decayed completely after some decay time $\tau_{c}$. After $\tau_{c}$ $f(\tau)$ grows linearly with $\tau$ and equation (\ref{ststep}) can be iterated which eventually results in the corresponding rate equations (\ref{rateeq}). 
%In principle, the HAM approach can be understood as some kind of iterated version of Fermi's Golden Rule.
Figure~\ref{dysonrandWW} and Fig.~\ref{dysonconstWW}
show a comparative calculation of $U_{1}(\tau)$ and $U_{2}(\tau)$ for both analyzed interactions. Whereas $U_{1}(\tau)$ is equal for both interactions, $U_{2}(\tau)$
will depend on the structure.

\begin{figure}[h!]
\begin{center}
\resizebox{0.47\columnwidth}{!}{
  \includegraphics{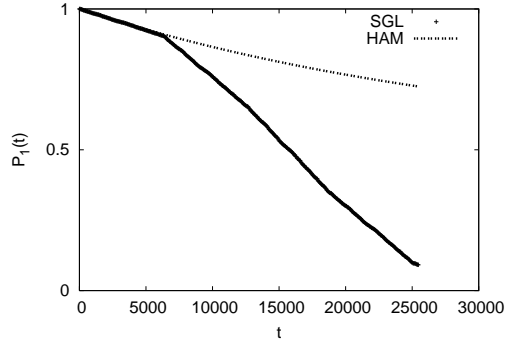} }
\caption{Dynamics of $P_{1}(t)$ for a constant interaction with $V_{ij}=1$, parameters:
 $N=500$, $\delta\epsilon=0.5$, $\lambda=0.00005$.}
\label{lesslambda}
\end{center}
\end{figure}

$U_{1}(\tau)$ and $U_{2}(\tau)$ must be compared in the time regime where equation (\ref{ststep}) is iterated, i.e., in the region after $\tau_{c}$ (here $\tau_{c}\approx 25$). This turns out to be equivalent to the beginning of the linear regime of $U_{1}(\tau)$. 
For the completely random interaction one finds that
$U_{1}(\tau)\gg U_{2}(\tau)$ at time $\tau_{c}$, so the second order truncation seems to be suggestive in this case.
%, although it is possible that the higher orders are not small compared to $U_{1}(\tau)$. 
There is no reason why higher orders beyond the second should have a decisive influence, since the second order already leads to a very good approximation.
This corresponds to the numerical calculations that showed statistical relaxation.
For the constant interaction $U_{2}(\tau)$ is already larger than $U_{1}(\tau)$ at time $\tau_{c}$. So the truncation is not justified which conforms to the numerically calculated non-statistical dynamics.
%supports the result that the numerically calculated dynamics were not statistical.
It should be remarked that this breakdown of the exponential behaviour is actually a problem of the structure of the interaction. It cannot be solved by simply
decreasing the average interaction strength $\lambda$. This is demonstrated in Fig.~\ref{lesslambda}.
$\lambda$ is chosen too small to fulfill criteria (\ref{kriterium2}). One again finds no complete decay into equilibrium here. However, one can see that Fermi's Golden Rule is fulfilled because of the good agreement at the very beginning. The deviation at some certain time nearly coincides with
the Heisenberg time that is equivalent to the recurrence time of the autocorrelation function of $V$ in our model with equidistant local energies ($\tau_{H}\approx 6200$). At this time the assumption of a completely decayed autocorrelation function necessarily becomes wrong and therefore the rate can no longer be considered as constant. Since the relaxation time of $P_{1}(t)$ is increased by a smaller $\lambda$, while the Heisenberg time is not, this finite size effect becomes important.

Generally, the full evaluation of the second order Dyson term or even higher orders is numerically and analytically very extensive. Instead of that one can
introduce a specific structural requirement for the interaction to estimate those contributions. In \cite{vanhove} Van Hove used a similar interaction structure to derive the possible occurrence of statistical relaxation. Basically, this would mean that $V$ features this so called Van Hove-structure, if $V^{2}$ is dominated by its diagonal elements in some sense. If the interaction does not have Van Hove-structure, then statistical relaxation cannot emerge. If the interaction
possesses Van Hove-structure, the dynamics are possibly but not inevitably statistical. So it can be regarded as a necessary criteria. One finds that the random interaction fulfills the Van Hove-structure, whereas the constant interaction does not. In this sense the resulting dynamics for both examples are explained correctly. In principle, the Van Hove-structure gives an estimation for the proper convergence of the respectively used perturbation expansion. It 
indicates that the higher orders should be small compared to the leading one, at least for times in the order of the relaxation time. 
%For the random and the constant interaction these considerations explain the resulting dynamics correctly.

\section{Analytic derivation of the solution of the rate equation}
\label{sec:2}
This Section is devoted to the derivation of the solution of the probability for the excitation to be on the left $P_{1}(t)$ or right hand side $P_{2}(t)$ for the case where the interaction is modeled by complex Gaussian entries in $V$. We will not show the total derivation as some intermediary results are too long but rather try to explain the main steps. This is why some results are just stated. What we mean by solution is the following. If we keep the interaction $V$ as a random matrix with its probability distribution instead of fixing it, then the time evolution operator, $e^{-i\left(H_{0}+\lambda V\right)t}$, becomes a random evolution operator. Thus $P_{1}(t)$ also contains this random character. We can then calculate its average, $\mathbf{E}[P_{1}(t)]=\int p(V) dV P_{1}(t)$, where $p(V)$ is the probability distribution over the random interaction. $p(V)$ is the product of the probability distribution of the complex Gaussian elements which can be written in a more compact form as $p(V)=\frac{1}{Z}e^{-\frac{N}{\sigma} \rm Tr \{V^{2}\}}$, where $Z$ is the partition function. We will calculate this average in the limit where the number of energy levels per subunit (there are 2 subunits in the present case) $N$ tends to $\infty$ and in the Van Hove limit. The Van Hove limit is given by $\lim_{\lambda\rightarrow 0}\lim_{t \rightarrow \infty} \lambda^2 t =T$. We call $T$ the macroscopic time. This limit represents some long time-weak coupling limit.
We will actually not calculate $\mathbf{E}[P_{1}(t)]$ directly. We will calculate $\mathbf{E}[P_{1}(t)-P_{2}(t)]$. Since we know that $P_{1}(t)+P_{2}(t)=1$ always holds (may it be averaged or not), we can obtain $\mathbf{E}[P_{1}(t)]$ from it.

The main idea of the proof and the below presented expansion is that, in the limits considered, the interference effects vanish. For this graphs are introduced which represent pairs of histories of the state of the system. The vanishing of interference effects means that for two different pairs of histories the contribution is much smaller than for equal or similar pairs. In some sense the quantum feature of interference becomes negligible when the limits are taken, thus making it more classical. It is basically this fact that allows for an autonomous equation to exist. For a pair of histories, or a graph, it is the random interaction, after averaging, that will decide on the weight given to it. It will also decide which class of pairs of histories, or graphs, are preponderant, thus contributing the most. These will be called simple graphs.
We start by writing down the observable we wish to consider and inserting the expansion of the time evolution operator.
\begin{eqnarray}
P_{1}(t)&=&\langle \psi_{0}|e^{iHt}\sum_{l=1}^{N}|1,l\rangle \langle1,l|e^{-iHt}|\psi_{0}\rangle \\
P^{l}_{1}(t)&=&\langle \psi_{0}|e^{iHt}|1,l\rangle \langle1,l|e^{-iHt}|\psi_{0}\rangle 
\end{eqnarray}
Every index $l$ represents the dependency on $E_{l}$ and $|\mu,l\rangle$ is a basis ket for the state being in unit $\mu$ on energy level $E_{l}$. \\
Loosely speaking the idea is to expand the time evolution operator in powers of the interaction. We have then powers of the random matrix over which we can average. Some parts won't contribute in the limit $N\rightarrow \infty$ and others won't contribute in the Van Hove limit. In the limit $N\rightarrow \infty$ we will keep our local spectrum bounded and so the energy level variables, $E_{l}$, will turn into continuous variables, $E$. $P^{l}_{1}(t)$ will become then $P_{1}(E,t)$.
We expand the evolution operator as
\begin{eqnarray}
e^{-iHt}&=&\sum_{n}(-i\lambda)^{n} \Gamma_{n}(t)\ , \\
\Gamma_{n}(t)&=& \int_{0}^{t}\dots \int_{0}^{t}ds_{0}\dots ds_{n} e^{-iH_{0}s_{0}}V\dots Ve^{-iH_{0}s_{n}} \delta\left(t-\sum_{j=0}^{n} s_{j} \right)\ .
\label{eq:gamma}
\end{eqnarray}
We thus have
\begin{eqnarray}
P^{l}_{1}(t)&=&\sum_{n,m}(i\lambda)^{m}(-i\lambda)^{n}\langle \psi_{0}|\Gamma_{m}^{\dagger}(t)|1,l\rangle \langle1,l|\Gamma_{n}(t)|\psi_{0}\rangle\ , \\
P^{l}_{1}(t)&=&\sum_{n,m}\sum_{l_{n},l'_{m}=1}^{N}\sum_{p_{n},p'_{m}=1}^{2}
\psi_{0}^{*}(p'_{m},l'_{m})\psi_{0}(p_{n},l_{n})i^{m}(-i)^{n}\lambda^{n+m}
\langle p'_{m},l'_{m}|\Gamma_{m}^{\dagger}|1,l\rangle \langle 1,l|\Gamma_{n}|p_{n},l_{n}\rangle \nonumber \ .\\
&&
\label{eq:P1}
\end{eqnarray}
%\begin{eqnarray}
%P^{l}_{1}(t)&=&\sum_{n,m}\sum_{l_{n},l'_{m}=1}^{N}\sum_{p_{n},p'_{m}=1}^{2}
%\psi_{0}^{*}(p'_{m},l'_{m})\psi_{0}(p_{n},l_{n})i^{m}(-i)^{n}\lambda^{n+m}
%\langle p'_{m},l'_{m}|\Gamma_{m}^{\dagger}|1,l\rangle \langle 1,l|\Gamma_{n}|p_{n},l_{n}\rangle 
%\label{eq:P1}\\
%P^{l}_{2}(t)&=&\sum_{n,m}\sum_{l_{n},l'_{m}=1}^{N}\sum_{p_{n},p'_{m}=1}^{2}
%\psi_{0}^{*}(p'_{m},l'_{m})\psi_{0}(p_{n},l_{n})i^{m}(-i)^{n}\lambda^{n+m} 
%\langle p'_{m},l'_{m}|\Gamma_{m}^{\dagger}|2,l\rangle \langle 2,l|\Gamma_{n}|p_{n},l_{n}\rangle
%\label{eq:P2}
%\end{eqnarray}
%%
We notice that if $n+m$ is odd, $\langle p'_{m},l'_{m}|\Gamma_{m}^{\dagger}|1,l\rangle \langle 1,l|\Gamma_{n}|p_{n},l_{n}\rangle$ is proportional to an odd number of random variables with average zero. The average of an odd number of random variables centered around zero is zero and so all terms with odd $n+m$ don't contribute to the average. For $n+m$ to be even we need both to be even or both to be odd.
The variables $p'_{m}$ and $p_{n}$ here stand for the unit $1$ or $2$. Since the interaction matrix makes an excitation hop from unit $1$ to unit $2$ and vice versa, we notice that if $n$ is even, then the product $\langle 1,l|\Gamma_{n}|p_{n},l_{n}\rangle$ is zero if $p_{n}$ is equal to $2$. Thus $p_{n}$ has to be equal to $1$. For $n$ odd we will only have a contribution for $p_{n}=2$.
We make this explicit in formula (\ref{eq:P1again}).
\begin{eqnarray}
P^{l}_{1}(t)&=&\sum_{n,m} \lambda^{2(n+m)}\sum_{l_{2n},l'_{2m}=1}^{N}
\psi_{0}^{*}(1,l'_{2m})\psi_{0}(1,l_{2n})i^{2m}(-i)^{2n} 
  \langle 1,l'_{2m}|\Gamma_{2m}^{\dagger}|1,l\rangle \langle 1,l|\Gamma_{2n}|1,l_{2n}\rangle \nonumber \\
&+&\sum_{n,m} \lambda^{2(n+m)+2}\sum_{l_{2n+1},l'_{2m+1}=1}^{N}
 \psi_{0}^{*}(2,l'_{2m+1})\psi_{0}(2,l_{2n+1})i^{2m+1}(-i)^{2n+1} \times \nonumber \\
&  &\langle 2,l'_{2m+1}|\Gamma_{2m+1}^{\dagger}|1,l\rangle \langle 1,l|\Gamma_{2n+1}|2,l_{2n+1}\rangle  \label{eq:P1again} \\
&=& F^{l}_{1}(t)+F^{l}_{2}(t)
\label{eq:F}
\end{eqnarray}
Notice that the first contribution is related to the initial state of unit $1$ and the second to the unit $2$.
%%
%\begin{eqnarray}
%P^{l}_{2}(t)&=&\sum_{n,m}\lambda^{2(n+m)} \sum_{l_{2n},l'_{2m}=1}^{N}
%\psi_{0}^{*}(2,l'_{2m})\psi_{0}(2,l_{2n})i^{2m}(-i)^{2n}
%\langle 2,l'_{2m}|\Gamma_{2m}^{\dagger}|2,l\rangle \langle 2,l|\Gamma_{2n}|2,l_{2n}\rangle \\
%&+& \sum_{n,m}\lambda^{2(n+m)+2} \sum_{l_{2n+1},l'_{2m+1}=1}^{N}
%\psi_{0}^{*}(1,l'_{2m+1})\psi_{0}(1,l_{2n+1})i^{2m+1}(-i)^{2n+1}
%\langle 1,l'_{2m+1}|\Gamma_{2m+1}^{\dagger}|2,l\rangle \langle 2,l|\Gamma_{2n+1}|1,l_{2n+1}\rangle \\
%&=& G^{l}_{2}(t)+G^{l}_{1}(t)
%\end{eqnarray}
%%
Of course, we have a similar expression for $P^{l}_{2}(t)$,
\begin{eqnarray}
P^{l}_{2}(t)&=& G^{l}_{2}(t)+G^{l}_{1}(t)\ .
\end{eqnarray}
When subtracting $P^{l}_{2}$ from $P^{l}_{1}$ we can group terms with the same initial data dependence that is $F^{l}_{1}$ with $G^{l}_{1}$ , and $F^{l}_{2}$ with $G^{l}_{2}$. We then have
\begin{equation}
\mathbf{E}[P^{l}_{1}(t)-P^{l}_{2}(t)]= \mathbf{E}[F^{l}_{1}(t)-G^{l}_{1}(t)] +\mathbf{E}[F^{l}_{2}(t)-G^{l}_{2}(t)]\ .
\label{eq:FG}
\end{equation}
We will thus compute one of these, the other being analogue.
%%
%%
%We now focus on terms such as $\langle 1,l'_{2m}|\Gamma_{2m}^{\dagger}|1,l\rangle \langle 1,l|\Gamma_{2n}|1,l_{2n}\rangle$.
Inserting (\ref{eq:gamma}) in $\langle 1,l|\Gamma_{n}|p_{n},l_{n}\rangle$ and identities after each interaction term we have 
\begin{eqnarray}
(-i)^{n}\langle 1,l|\Gamma_{n}|p_{n},l_{n}\rangle &=& \int_{0}^{t}\dots \int_{0}^{t}ds_{0}\dots ds_{n} e^{-iE_{l}s_{0}}e^{-iE_{l_{1}}s_{1}}\dots e^{-iE_{l_{n}}s_{n}} \delta\left(t-\sum_{j=0}^{n} s_{j} \right) \\
&\times& (-i)^{n}\langle 1,l|V|p_{1},l_{1}\rangle \langle p_{1},l_{1}|V|p_{2},l_{2}\rangle \dots \langle p_{n-1},l_{n-1}|V|p_{n},l_{n}\rangle\ . \nonumber
\end{eqnarray}
For a shorter notation we define
\begin{eqnarray}
K^{n}(t,\{E_{l_{j}}\}) &=& (-i)^{n}\int_{0}^{t}\dots \int_{0}^{t}ds_{0}\dots ds_{n} e^{-iE_{l}s_{0}}e^{-iE_{l_{1}}s_{1}}\dots e^{-iE_{l_{n}}s_{n}} \delta\left(t-\sum_{j=0}^{n} s_{j} \right), 
\label{eq:K} \\
L^{n}(\{l_{j}\},\{p_{i}\})&=& \langle 1,l|V|p_{1},l_{1}\rangle \langle p_{1},l_{1}|V|p_{2},l_{2}\rangle \dots \langle p_{n-1},l_{n-1}|V|p_{n},l_{n}\rangle\ . \label{eq:L}
\end{eqnarray}
%%Anothe usefull representation of Eq. (\ref{eq:K}) is as follows. \\
%%Notice first that the integration over the $s_{j}$ variables can be extended from $t$ untill $\infty$ since the $\delta$ function ensures $s_{j}<t$.
%%We choose the following representation of the delta function:
%%\begin{equation}
%%\int_{-\infty}^{\infty}d\alpha e^{-i\alpha \left(t-\sum_{j=0}^{n}s_{j}\right)}e^{\eta\left(t-\sum_{j=0}^{n}s_{j}\right)}
%%\end{equation}
%%
%%The first exponential ensures the $\delta $ function, while the second is an identity inserted for convenience.
%%Grouping terms multiplied by $s_{j}$ in the exponential we have then terms such as 
%%
%%\begin{equation}
%%\int_{0}^{\infty}ds_{j}e^{-is_{j}\left(E_{j}-\alpha -i\eta \right)}=\frac{-i}{E_{j}-\alpha -i\eta}
%%\end{equation} 
%%
%%\begin{equation}
%%K^{n}(t,\{E_{j}\})=i\int_{-\infty}^{\infty}d\alpha e^{-i\alpha t}e^{\eta t}\frac{-1}{E_{l}-\alpha-i\eta} \frac{-1}{E_{l_{1}}-\alpha-i\eta} \dots %%\frac{-1}{E_{l_{n}}-\alpha-i\eta} \label{eq:K2}
%%\end{equation}

All of the randomness is encoded in $L^{n}(\{l_{j}\},\{p_{i}\})$ and we know the $\{p_{i}\}$ variables are determined if $n$ is odd or even.
With this notation we have for the first term in Eq. (\ref{eq:P1again})
\begin{eqnarray}
& &\sum_{l_{2n},l'_{2m}=1}^{N}
\psi_{0}^{*}(1,l'_{2m})\psi_{0}(1,l_{2n})i^{2m}(-i)^{2n} 
 \langle 1,l'_{2m}|\Gamma_{2m}^{\dagger}|1,l\rangle \langle 1,l|\Gamma_{2n}|1,l_{2n}\rangle \nonumber \\
&=&\sum_{\{l_{i},l'_{j}\}}\psi_{0}^{*}(1,l'_{2m})\psi_{0}(1,l_{2n})K^{2n}(t,\{E_{l_{i}}\})\bar{K}^{2m}(t,\{E_{l'_{i}}\})
L^{2n}(\{l_{j}\},\{p_{j}\}) \bar{L}^{2m}(\{l'_{j}\},\{p'_{j}\}) .\label{eq:firstterm}
\label{eq:doublegamma}
\end{eqnarray}
To average we must average over $L^{n}\bar{L}^{m}$. This term is a product of Gaussian complex variables. It only contributes if these variables correlate. This will introduce relations between the different indices $\{l_{i},l'_{j},l\}$. According to Wigner's theorem we have the following formula to calculate the average over a product of Gaussian random variables.
\begin{equation}
\mathbf{E}[X_{1} \dots X_{2n} ]=\sum_{C_{\pi}}\prod_{(i,j)\in C_{\pi}}\mathbf{E}[X_{i}X_{j}]
\end{equation}  
Here $C_{\pi}$ is a permutation of the set of indices $\{1,\dots 2n\}$. In the present case each $X_{j}$ corresponds to matrix elements of the type $\langle p_{1},l_{1}|V|p_{2},l_{2}\rangle$. The pairings induce a graph structure on $L^{n}\bar{L}^{m}$. Pictorially an example is given in Fig. $1$. 
\begin{figure}[h]
\begin{center}
% Use the relevant command for your figure-insertion program
% to insert the figure file.
% For example, with the option graphics use
\resizebox{0.99\columnwidth}{!}{
  \includegraphics{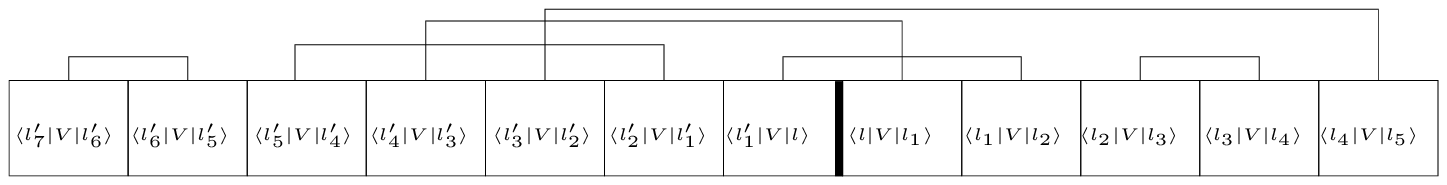}}
\end{center}
\caption{Example of graph representation for a pairing with $n=5$ and $m=7$. 
This graph represents the following contribution: \newline
 $\mathbf{E}[\langle l'_{7}|V|l'_{6}\rangle \langle l'_{6}|V|l'_{5}\rangle] \times
\mathbf{E}[\langle l'_{5}|V|l'_{4}\rangle \langle l'_{2}|V|l'_{1}\rangle]\times
%\mathbf{E}[\langle l'_{4}|H_{I}|l'_{3}\rangle \langle l|H_{I}|l_{1}\rangle] 
%\mathbf{E}[\langle l'_{3}|H_{I}|l'_{2}\rangle \langle l_{4}|H_{I}|l_{5}\rangle]
\dots
\mathbf{E}[\langle l'_{1}|V|l \rangle \langle l_{1}|V|l_{2}\rangle]\times
\mathbf{E}[\langle l_{2}|V|l_{3}\rangle \langle l_{3}|V|l_{4}\rangle]
$. \newline 
All $p_{j} $ indices are omitted for lighter notation.}%Please write your figure caption here.}
\label{fig:1}       % Give a unique label
\end{figure}
The average of one pairing is given in Eq. (\ref{eq:onepair}).
\begin{equation}
\mathbf{E}[\langle 1,l_{i}|V|2,l_{i+1}\rangle \langle 2,l_{j}|V|1,l_{j+1}\rangle]
=\frac{\sigma}{N}\delta_{l_{i},l_{j+1}} \delta_{l_{i+1},l_{j}} \label{eq:onepair}
\end{equation}
Notice that the averaging enforces identities amongst $\{l_{i},{l_{i+1},l_{j},l_{j+1}}\}$ and hence amongst the energies $\{E_{l_{i}},E_{l_{i+1}}, E_{l_{j}},E_{l_{j+1}}\}$ in Eq. (\ref{eq:doublegamma}).
The question is then how much graphs weight and what kinds of graphs are important. We focus on the weight they have.

Over many pairings the delta functions are responsible for the graph structure. However, we notice the weight of every pair will always be the same. Thus we conclude the weight of a graph on $L^{2n}\bar{L}^{2m}$ in (\ref{eq:doublegamma}) is given by $\left(\frac{\sigma}{N}\right)^{n+m}$.
$\mathbf{E}[L^{2n}(\{l_{j}\},\{p_{j}\}) \bar{L}^{2m}(\{l'_{j}\},\{p'_{j}\})]$ is represented by a graph and enforces identities amongst the energies $\{E_{l_{i}}, E_{l'_{i}}\}$.
We can then rewrite (\ref{eq:doublegamma}) as
\begin{equation}
\sum_{\textrm{\tiny graphs } C_{\pi}(2n,2m)}\left(\sum_{\textrm{\tiny indep.}\{l_{i},l'_{j}\}}\left(\frac{\sigma}{N}\right)^{n+m}\psi_{0}^{*}(1,l'_{2m})\psi_{0}(1,l_{2n})K^{2n}_{\pi}(t,\{E_{l_{i}},E'_{l_{i}}\})\bar{K}^{2m}_{\pi}(t,\{E_{l_{i}},E_{l'_{i}}\})\right). \nonumber
\end{equation}
%%
%\bar{K}^{2m}_{\pi}(t,\{E_{l_{i}},E_{l'_{i}}\})
Here $K^{2n}_{\pi}(t,\{E_{l_{i}},E'_{l_{i}}\})\bar{K}^{2m}_{\pi}(t,\{E_{l_{i}},E_{l'_{i}}\})$
stands for the function $K^{2n}\bar{K}^{2m}$ from Eq.(\ref{eq:doublegamma}) but with the identities amongst the energy variables imposed.
It can be shown that for $C_{\pi}(2n,2m)$ graphs the maximum of independent $\{l_{i},l'_{j}\}$ is $n+m$ and that we only have this for a certain class of graphs, which we will call simple graphs (S. graphs). For this class of graphs holds that the identity $l_{2n}=l'_{2m}$ is fulfilled. The fact that this identity is fulfilled implies that  $\psi_{0}^{*}(1,l'_{2m})\psi_{0}(1,l_{2n})=P_{1}^{l_{2n}}(t=0)$ .
%Notice however that in (\ref{eq:doublegamma}) we are summing over almost all  $\{l_{i},l'_{j}\}$. If Eq. (\ref{eq:doublegamma}) is inserted in  Eq. (\ref{eq:P1}) we have a sum over all $\{l_{i},l'_{j}\}$. These sums, which are there for any graph and run from $1$ untill $N$ , are those that will compensate $\left(\frac{1}{N}\right)^{n+m}$ factor. 

\begin{figure}[h]
\begin{center}
% Use the relevant command for your figure-insertion program
% to insert the figure file.
% For example, with the option graphics use
\resizebox{0.99\columnwidth}{!}{
  \includegraphics{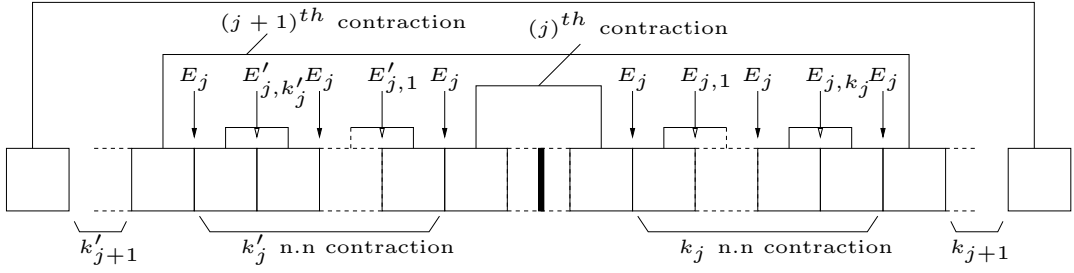}}
\end{center}
\caption{Sketch of simple graphs}.
\label{fig:2}       % Give a unique label
\end{figure}

%In Eq. (\ref{eq:doublegamma}), for example, there are in total $2n+2m$  $\{l_{i},l'_{j}\}$ variables. If we roughly say every $l$ contributes $N$ times in the sum this would give us a factor of $N^{2n+2m}$. But remember that the graph imposes relations between the $\{l_{i},l'_{j}\}$ variables and so it is the number of independent variables, after imposing the delta relations, that decide on the contribution for a certain graph. It turns out that the maximum number of independent variables one can have from the set $\{l_{i},l'_{j}\}$, in Eq. (\ref{eq:doublegamma}), is just $n+m$.
%So effectivly we are summing over $n+m$ variables from $1$ to $N$ wheighted by a $\left(\frac{\sigma}{N}\right)^{n+m}$ factor. 
We would then have
\begin{equation}
\sum_{w_1}\dots \sum_{w_{n+m}}\left(\frac{\sigma}{N}\right)^{n+m} \rightarrow \sigma^{n+m}\prod_{j=1}^{n+m}\int dE_{j}=\sigma^{n+m}\int dE\ .
\label{eq:pictorialsum}
\end{equation}
The $w_{j}$ here represent the independent $l_{j}$ and $\int dE$ is a short notation for the multiple integrals.\\
The sums over independent variables turn into integrals over, what has become, continuous energy variables $E_{j}$. The limit as $N\rightarrow \infty$ of the average of Eq.(\ref{eq:firstterm}) then becomes  
\begin{eqnarray}
&&\mathbf{E}[F_{1}(E_{0},t)]=\lim_{N\rightarrow \infty}\mathbf{E}[F_{1}^{l}(t)] \label{eq:limF}\\
&&=\sum_{n,m} (\lambda^2)^{n+m}\sum_{\textrm{\tiny S graphs } C_{\pi}(2n,2m)}\sigma^{n+m}\left(\int dE P_{1}(E_{2n},t=0)K^{2n}_{\pi}(t,\{E_{i}\})\bar{K}^{2m}_{\pi}(t,\{E_{j}\})\right). \nonumber
\end{eqnarray}
$E_{0}$ is the now continuous variable $E_{l}$. 
We now explain what type of graphs contribute in the way previously described. Among all graphs the ones of the type shown in Fig. \ref{fig:2} are simple graphs.
Figure \ref{fig:2} represents graphs where $\bar{n}$ random variables from the left hand side are paired with one on the right hand side, and in between each pair of this left-right type of pairs we have a certain amount of pairings with their nearest neighboring random variables. That is, in between the $j^{th}$ pairing from left to right and the $(j+1)^{th}$ pairing from left to right we have $k'_{j}$ pairings on the left and $k_{j}$ pairings on the right.
It is very important to notice that the information of the graph is now coded in the set of number $\{\bar{n},k'_{j},k_{j}\}$ which determine uniquely the graph.
%\begin{eqnarray}
%F^{l}_{1}(t)&=&
%\sum_{n,m}\lambda^{2(n+m)} \sum_{l_{i},l'_{j}=1}^{N}
%\psi_{0}^{*}(1,l'_{2m})\psi_{0}(1,l_{2n})i^{2m}(-i)^{2n}
%K^{2n}(t,\{E_{l_{i}}\})\bar{K}^{2m}(t,\{E_{l'_{i}}\}) \nonumber \\
%&\times& \sum_{C_{\pi}} \left(\frac{\sigma}{N}\right)^{n+m} \tilde{C}_{\pi}(n,m,\{l_{i},l'_{j},l\})
%\end{eqnarray}
%%
For such graphs the exponentials of the $K^{n}\bar{K}^{m}$ factors in Eq. (\ref{eq:doublegamma}) and Eq. (\ref{eq:limF})  in between the $j^{th}$ and $(j+1)^{th}$ contraction have the form
%By use of Eq. (\ref{eq:K2}) and (\ref{eq:doublegamma}) the part in between the $j^{th}$ and $(j+1)^{th}$ contraction of one graph with set of indices $\{\bar{n},k'_{j},k_{j}\}$ is then
%%
\begin{eqnarray}
C_{j}=(-i)^{k_{j}+1} e^{-iE_{j}\sum_{l=1}^{k_{j}+1}s^{j}_{l} } \left(\prod_{q=1}^{k_{j}}-ie^{-iE_{j,q}s_{jq}}\right) (i)^{k'_{j}+1} e^{iE_{j}\sum_{l=1}^{k_{j}+1}\tau^{j}_{l}}\left(\prod_{q'=1}^{k'_{j}}ie^{iE'_{j,q'}\tau_{jq'}}\right).
\end{eqnarray}
$C_{j}$ depends on many variables ( $C_{j}=C_{j}(E_{j},E_{j,q},E'_{j,q'},s_{jq},\tau_{jq'},s^{j}_{l},\tau^{j}_{l})$ ) which we didn't write explicitly.
In order to calculate the contribution of a graph we need to integrate the product of all $C_{j}$ functions over all of these variables respecting the delta functions of Eq. (\ref{eq:K}). It can be shown that in the Van Hove limit each integral over the functions depending on $E_{j,q}$, $s_{jq}$, $E'_{j,q'}$ and $\tau_{jq'}$ can be replaced by a given constant.
We have then
\begin{eqnarray}
C_{j}(E_{j},s^{j}_{l},\tau^{j}_{l})=(-i)^{k_{j}+1} e^{-iE_{j}\sum_{l=1}^{k_{j}+1}s^{j}_{l} }\Theta^{k_{j}} (i)^{k'_{j}+1} e^{iE_{j}\sum_{l=1}^{k_{j}+1}\tau^{j}_{l}}\bar{\Theta}^{k'_{j}}
\end{eqnarray}
and
\begin{eqnarray}
& &Q(C_{\pi}(\bar{n},\{k_{i},k'_{j}\}),E_{0},t)=\left(\int dE P_{1}(E_{2n},t=0)K^{2n}_{\pi}(t,\{E_{i}\})\bar{K}^{2m}_{\pi}(t,\{E_{j}\})\right) \label{eq:Q}\\
& &=\prod_{j=1}^{\bar{n}}\int dE_{j}P_{1}(E_{\bar{n}},0)\prod_{l=1}^{k_{j}+1} \int_{0}^{\infty}ds^{j}_{l} \prod_{l=1}^{k'_{j}+1} \int_{0}^{\infty}d\tau^{j}_{l}
\delta\left(t-\sum_{j,l}s^{j}_{l}\right)\delta\left(t-\sum_{j,l}\tau^{j}_{l}\right)C_{j}(E_{j},s^{j}_{l},\tau^{j}_{l}).
\nonumber
\end{eqnarray}
For the integration over the time variables we have the identity (\ref{eq:tint}).
\begin{eqnarray}
 \prod_{j=1}^{\bar{n}} \prod_{l=1}^{k_{j}+1} \int_{0}^{\infty}ds^{j}_{l} \delta\left(t-\sum_{j,l}s^{j}_{l}\right) =
 \prod_{j=1}^{\bar{n}} \int_{0}^{\infty} ds_{j} \delta\left(t-\sum_{j}s_{j}\right) \prod_{l=1}^{k_{j}+1}\int_{0}^{\infty}ds^{j}_{l} \delta\left(s_{j}-\sum_{l}s^{j}_{l}\right)
 \label{eq:tint} 
\end{eqnarray}
We have the same identity for the $\tau$ variables. We see that with this identity we can make the integrations over the $s^{j}_{l}$ variables and $\tau^{j}_{l}$ variables easily because the exponential in $C_{j}$ depends on $\sum_{l=1}^{k_{j}+1}s^{j}_{l}$ and $\sum_{l=1}^{k'_{j}+1}\tau^{j}_{l}$.
%We can thus perform the last integrations over the $s^{j}_{l}$ variables:
\begin{eqnarray}
\tilde{C_{j}}(E_{j},s_{j},\tau_{j},k_{j},k'_{j})&=& \prod_{l=1}^{k_{j}+1}\int_{0}^{\infty}ds^{j}_{l}\delta\left(s_{j}-\sum_{l}s^{j}_{l}\right) 
\prod_{l=1}^{k'_{j}+1}\int_{0}^{\infty}d\tau^{j}_{1}\delta\left(\tau_{j}-\sum_{l}\tau^{j}_{l}\right)
C_{j}(E_{j},s^{j}_{l},\tau^{j}_{l}) \nonumber \\
&=&(-i)^{k_{j}+1} (i)^{k'_{j}+1} e^{-iE_{j} s_{j} } \frac{(s_{j})^{k_{j}}}{k_{j}!} \Theta^{k_{j}}  e^{iE_{j}\tau_{j}} \frac{(\tau_{j})^{k'_{j}}}{k'_{j}!}\bar{\Theta}^{k'_{j}} \label{eq:Ctilde}
\end{eqnarray}
We then have by Eq. (\ref{eq:Q}), (\ref{eq:tint}) and (\ref{eq:Ctilde})
\begin{eqnarray}
& &Q(C_{\pi}(\bar{n},\{k_{i},k'_{j}\}),E_{0},t) \nonumber \\
&=&\prod_{i=1}^{\bar{n}}\int dE_{i} \prod_{j=0}^{\bar{n}}\int ds_{j}\int d\tau_{j} \tilde{C_{j}}(E_{j},s_{j},\tau_{j},k_{j},k'_{j}) \delta\left(t-\sum_{j}s_{j}\right)
\delta\left(t-\sum_{j}\tau_{j}\right).
\end{eqnarray}
$F_{1}^{l}(t)$ has become $F_{1}(E_{0},t)$ due to $E_{l}\rightarrow E_{0}$. Since our graphs are determined by the variables $\{\bar{n},k_{j},k'_{j}\}$ and we have to sum over all possible graphs, this sum over all graphs becomes a sum over all possible values of $\{\bar{n},k_{j},k'_{j}\}$. 
\begin{equation}
\mathbf{E}[F_{1}(E_{0},t)]=
\sum_{\textrm{\tiny all even }\bar{n}} (\sigma\lambda^2)^{\bar{n}} \sum_{\{k_{i},k'_{j}\}}
(\sigma\lambda^2)^{ k_{i}+k'_{j}}
Q(C_{\pi}(\bar{n},\{k_{i},k'_{j}\}),E_{0},t) \label{eq:EF}
\end{equation}
The fact that $\bar{n}$ is restricted to even numbers comes from Eq. (\ref{eq:F}). That means, $F_{1}(E_{0},t)$ is the contribution when we have an even number of interaction matrices on the left and on the right. This is guaranteed when $\bar{n}$ is even. $G_{1}(E_{0},t)$ turns out to be the sum over all odd $\bar{n}$. 
By grouping together all terms which are to the power of $k_{j}$ and $k'_{j}$ we can now sum over these.

\begin{eqnarray}
&&\sum_{k_{j},k'_{j}}(\sigma\lambda^2)^{k_{j}+k'_{j}} Q(C_{\pi}(\bar{n},\{k_{i},k'_{j}\}),E_{0},t) = \label{eq:QSum}\\ %\prod_{j=0}^{\bar{n}}\int dE_{j}P_{1}(E_{\bar{n}},0) \\
%&&\prod_{j=0}^{\bar{n}}\int_{0}^{\infty}ds_{j}\delta\left(t-\sum s_{j}\right)
%e^{-is_{j}\sigma\lambda^{2}\Theta}e^{-iE_{j}s_{j}} 
%\prod_{j={0}}^{\bar{n}}\int_{0}^{\infty}d\tau_{j}\delta\left(t-\sum \tau_{j}\right)
%e^{i\tau_{j}\sigma\lambda^{2}\bar{\Theta}}e^{iE_{j}\tau_{j}} \nonumber \\
&& e^{-it\sigma\lambda^{2}(\Theta-\bar{\Theta})}\prod_{i=1}^{\bar{n}}\int dE_{i}P_{1}(E_{\bar{n}},0)\prod_{j=0}^{\bar{n}}
\int_{0}^{\infty}ds_{j}d\tau_{j}\delta\left(t-\sum s_{j}\right)\delta\left(t-\sum \tau_{j}\right)
e^{-iE_{j}(s_{j}-\tau_{j})} \nonumber
\end{eqnarray}
In the first exponent $\Theta-\bar{\Theta}$ is thus the imaginary part of $\Theta$. We have not derived this constant but it turns out to have the following imaginary part.
\begin{equation}
\Theta-\bar{\Theta}= -i2\pi
\end{equation}
Thus
\begin{equation}
\lim_{\textrm{\tiny Van Hove}}e^{-it\sigma\lambda^{2}(\Theta-\bar{\Theta})}=e^{-2\pi\sigma T}\ .
\end{equation}
We now turn to the first part of Eq. (\ref{eq:FG}). By Eq. (\ref{eq:EF}) and Eq. (\ref{eq:QSum}) we have an expression for $\mathbf{E}[F_{1}(E,t)]$. A similar one can be obtained for $\mathbf{E}[G_{1}(E,t)]$. We group these together in the following equations.
\begin{eqnarray}
\mathbf{E}[F_{1}(E,t)-G_{1}(E,t)]&=&\sum_{ \textrm{\tiny all even }\bar{n}}(\sigma\lambda^2)^{\bar{n}}\sum_{\{k_{i},k'_{j}\}}
(\sigma\lambda^2)^{k_{i}+k'_{j}}Q(C_{\pi}(\bar{n},\{k_{i},k'_{j}\}),E,t)\\
&& -\sum_{ \textrm{\tiny all odd }\bar{n}}(\sigma\lambda^2)^{\bar{n}}\sum_{\{k_{i},k'_{j}\}}
(\sigma\lambda^2)^{k_{i}+k'_{j}}Q(C_{\pi}(\bar{n},\{k_{i},k'_{j}\}),E,t) \nonumber \\
&=&\sum_{ \textrm{\tiny all }\bar{n}}
e^{-it\sigma\lambda^{2}(\Theta-\bar{\Theta})}\prod_{i=1}^{\bar{n}}\int dE_{i}P_{1}(E_{\bar{n}},0)\prod_{j=0}^{\bar{n}} \label{eq:final}\\
&&(-\sigma\lambda^{2})^{\bar{n}}\int_{0}^{\infty}ds_{j}d\tau_{j}\delta\left(t-\sum s_{j}\right)\delta\left(t-\sum \tau_{j}\right)
e^{-iE_{j}(s_{j}-\tau_{j})} \nonumber
\end{eqnarray}
To calculate the integral part we introduce a change of variables, Eq. (\ref{eq:changeofvar}), in the last line of Eq. (\ref{eq:final}).
\begin{equation}
\alpha_{j}=\lambda^2\frac{s_{j}+\tau_{j}}{2},b_{j}=\frac{s_{j}-\tau_{j}}{2} \label{eq:changeofvar}
\end{equation}
\begin{equation}
(-\sigma)^{\bar{n}}\prod_{j=0}^{\bar{n}}\int_{0}^{T} d\alpha_{j}\delta\left(T-\sum_{0}^{\bar{n}}\alpha_{j}\right)
\int_{\frac{-\alpha_{j}}{\lambda^2}}^{\frac{+\alpha_{j}}{\lambda^2}}
db_{j}\delta\left(\sum_{j=0}^{\bar{n}}b_{j}\right)e^{-i\sum_{j=0}^{\bar{n}}b_{j}E_{j}}
\label{eq:changeab}
\end{equation}
In the Van Hove limit Eq. (\ref{eq:changeab}) turns into
\begin{eqnarray}
(-\sigma)^{\bar{n}} \frac{T^{\bar{n}}}{\bar{n}!}\prod_{j=0}^{\bar{n}-1}2\pi \delta\left(E_{j}-E_{\bar{n}}\right) \ .\label{eq:limitofchangeofvar}
\end{eqnarray}
Inserting Eq. (\ref{eq:limitofchangeofvar}) in Eq. (\ref{eq:final}) we obtain
\begin{eqnarray}
& &\lim_{\textrm{\tiny Van Hove}}\int dE \left(\mathbf{E}[F_{1}(E,t)-G_{1}(E,t)]\right) \nonumber \\%% = F_{1}(T)-G_{1}(T) \\
& = & \sum_{ \textrm{\tiny all }\bar{n} } e^{-2\pi \sigma T}
\prod_{j=1}^{\bar{n}}
\int dE_{j}P_{1}(E_{\bar{n}},0)(-\sigma)^{\bar{n}} \frac{T^{\bar{n}}}{\bar{n}!}\prod_{j=0}^{\bar{n}-1}2\pi \delta\left(E_{j}-E_{\bar{n}}\right) \nonumber \\
&=& \sum_{ \textrm{\tiny all }\bar{n} } e^{-2\pi \sigma T}
\int dE P_{1}(E,0)\frac{(-2\sigma\pi T)^{\bar{n}}}{\bar{n}!} \nonumber \\
&=& e^{-4\pi \sigma T} \int dE P_{1}(E,0) \ .\label{eq:Pdecay}
\end{eqnarray}
In the same manner we can calculate  $\int dE \left(\mathbf{E}[F_{2}(E,t)-G_{2}(E,t)]\right)$ in the Van Hove limit which will have the same form as Eq.(\ref{eq:Pdecay}) but will depend on $P_{2}(E,0)$ instead of $P_{1}(E,0)$.
Using these last results in Eq. (\ref{eq:FG}) we finally have
\begin{equation}
\lim_{\textrm{\tiny Van Hove}} \lim_{N\rightarrow \infty} \left(\mathbf{E}[P_{1}(t)-P_{2}(t)]\right)= e^{-4\pi \sigma T} \left(P_{1}(0)- P_{2}(0) \right)\ . 
\end{equation} 
%
%Thus 
%%\begin{eqnarray}
%\lim_{\textrm{\tiny Van Hove}} \lim_{N\rightarrow \infty} \left(\mathbf{E}[P_{1}(t)]\right)&=& \frac{1+e^{-4\pi \sigma T} \left(P_{1}(0)- P_{2}(0) \right)}{2} \\
%\lim_{\textrm{\tiny Van Hove}} \lim_{N\rightarrow \infty} \left(\mathbf{E}[P_{1}(t)]\right)&=& \frac{1+e^{-4\pi \sigma T} \left(P_{2}(0)- P_{1}(0) \right)}{2}
%\end{eqnarray} 
These give us the solutions to the rate equations with a rate of $4\pi \sigma$. 

\section{Conclusion}

The analytical derivation has shown that the dynamics of the occupation probabilities of single subunits are statistical for the average over the whole ensemble of
possible interactions for $N\rightarrow\infty$ and in the Van Hove-limit. This means that if "most" members of the ensemble (a dense subset of the ensemble) give the same type of relaxation, then there is statistical relaxation which is then a typical feature for members of the ensemble.
The numerical calculations for a finite size version of the system demonstrate that the
occurrence of statistical relaxation actually depends on the structure of a concrete realization of the interaction. A random interaction reproduces the results for the average. Nevertheless, the example of the constant interaction indicates that there are exceptions as the dynamics are not statistical at all in this case. 
The fact that a fixed randomly chosen interaction reproduces the same type of relaxation as the average strongly supports that the statistical relaxation is a typical feature because it represents the majority of all possible interactions very well. 

%and \cite{RefJ}
%\subsection{Subsection title}
%\label{sec:2}
%as required. Don't forget to give each section
%and subsection a unique label (see Sect.~\ref{sec:1}).
%

%\begin{figure}
% Use the relevant command for your figure-insertion program
% to insert the figure file.
% For example, with the option graphics use
%\resizebox{0.75\columnwidth}{!}{%
%  \includegraphics{fig1.eps} }
%\caption{Please write your figure caption here.}
%\label{fig:1}       % Give a unique label
%\end{figure}
%
% For tables use
%\begin{table}
%\caption{Please write your table caption here.}
%\label{tab:1}       % Give a unique label
% For LaTeX tables use
%\begin{tabular}{lll}
%\hline\noalign{\smallskip}
%first & second & third  \\
%\noalign{\smallskip}\hline\noalign{\smallskip}
%number & number & number \\
%number & number & number \\
%\noalign{\smallskip}\hline
%\end{tabular}
%\end{table}
%

\end{document}